\documentstyle[11pt]{article}

\advance\textheight by \topskip
\newcommand{\be}{\begin{eqnarray}}
\newcommand{\ee}{\end{eqnarray}}
\newcommand{\bfl}{\begin{flushleft}}
\newcommand{\efl}{\end{flushleft}}
\newcommand\ie {{\it i.e. }}
\newcommand\eg {{\it e.g. }}

\newcommand\half{\frac 1 2 }

\begin{document}

\title{Generalized statistics and the algebra of observables\thanks{Invited
lecture given at the 4th International School of Theoretical Physics, {\em
Symmetry and Structural Properties of Condensed Matter}, Zaj\c{a}czkowo,
28.August -- 4.September 1996.}} 

\author{ Jon Magne Leinaas\\  Centre for Advanced Study, \\ Drammensveien 78,
\\N--0271 Oslo, Norway \\ and \\Department of Physics, University of Oslo,\\
P.O. Box 1048 Blindern,\\ N--0316 Oslo 3, Norway.\thanks{Permanent address.}}

\date{
}

\maketitle

\begin{abstract} A short review is given of how to apply the algebraic
Heisenberg quantization scheme to a system of identical particles. For two
particles in one dimension the approach leads to a generalization of the Bose
and Fermi description which can be expressed in the form of a $1/x^2$
statistics interaction between the particles. For an N-particle system it is
shown how a particular infinite-dimensional algebra arises as a generalization
of the
$su(1,1)$ algebra which is present for the two-particle system.    
\end{abstract}

\vspace* {-145 mm}
\begin{flushright}  Oslo-TP 15-96 \\
November-1996 \\
\end{flushright}
\vskip 135mm

\section{Introduction} 
In recent years the possibilities of introducing generalized types of particle
statistics have attracted a great deal of attention. Although there is no reason
to doubt the traditional classification of elementary particles into bosons and
fermions, the investigation of natural generalizations of these two classes of
particles is an interesting exercise from a theoretical point of view. On one
side the investigation shows in what way bosons and fermions are special. On
the other side such generalizations may be relevant for particles that are not
elementary. This has been demonstrated in the case of the (fractional) quantum
Hall effect, where the quasi-particle excitations are belived to have anyonic
properties. (For an introduction to anyons and generalized statistics,
see for example
\cite{Leinaas 92} in the proceedings of the summer school of 1992.)

There are several different approaches possible when discussing generalizations
of bosons and fermions. I will loosely divide them into three classes. The
first one may be referred to as the {\em quantum mechanics} approach. This
approach adresses the question of how to quantize a system of identical
particles in such a way that the indistinguishability of the particles is taken
care of in the proper way
\cite{Laidlaw 71,Leinaas 77}. It is well known that such an analysis leads most
naturally to consider {\em anyons} as a generalization of bosons and fermions in
two-dimensional systems \cite{Leinaas 77,Goldin 80,Wilczek 82}. The second one
is the {\em statistical mechanics} approach. In this approach one considers
generalizations of the statistical distributions associated with bosons and
fermions
\cite{Haldane 91,Isakov 94,Wu 94}. Thus, one considers generalizations of the
Pauli exclusion principle in the form of modified rules for filling single
particle levels. The third one is the {\em second quantization} approach. Here
one considers the question of possible generalizations of the commutation and
anti-commutation relations between field operators in the cases of bosons and
fermions respectively \cite{Green 53,Greenberg 90,Ohnuki 82}. These three
approaches are not equivalent, although in some cases they lead to the same
type of generalizations. I will here only discuss the quantum mechanics
approach.

There are in fact (at least) two different ways, which both seem natural, to
approach the question of how to properly quantize a system of identical
particles. The first one is the {\em Schr\"odinger} approach, which leads to
the concept of anyons. In this approach one focuses the attention on the
configuration space of the system and introduces wave functions which are
defined on this space \cite{Leinaas 77}.  What is special about the
configuration space for a system of identical particles is that the
$N$-particle space is not the product space of $N$ single particle spaces.
Instead the correct configuration space is derived from the product space by
introducing identifications between points which correspond to reordering of
the particles. The result is a space with singularities, and there is a phase
factor associated with the wave functions when a singularity is encircled.
Such an encircling is physically associated with the interchange of the
position of two particles, and the corresponding phase factor is $+1$ for
bosons and $-1$ for fermions. In two space dimensions other phase factors are
conceivable and intermediate (anyonic) types of statistics is a possibility.
(The path integral formulation may  be regarded as a distinct approach, but
also this is based on the correct description of the configuration space, and
it essentially gives the same result as the Schr{\"o}dinger quantization
\cite{Laidlaw 71,Wu 84}.) 

The second approach ({\em Heisenberg} quantization) focuses the attention on the
observables rather than the wave functions. Quantization is introduced in the
form of a set of fundamental commutation relations between observables. Also
here the indistinguishability of the particles interferes in a fundamental way
with the quantization of the system. The main point is that the observables are
symmetric with respect to the interchange of particles and the fundamental
commutation relations are changed as compared with a system of non-identical
particles, where the observables are non-symmetric. As a consequence of this,
new possibilities appear and these are interpreted as generalized statistics. A
somewhat surprising result is that the Heisenberg and Schr\"odinger approaches
give rise to non-equivalent generalizations of  bosons and fermions
\cite{Leinaas 88,Leinaas 91}. 

It is the algebraic Heisenberg quantization scheme which will be discussed here.
Quantization is formulated as a program for finding irreducible representations
of a fundamental Lie algebra of observables. The program will first be
illustrated by two simple examples and the system of two identical particles in
one dimension will then discussed in some detail. Different types of statistics
will be shown to correspond to inequivalent representations of the same
fundamental algebra. It is argued that the {\em Calogero} model \cite{Calogero
69}, \ie a system of paticles with
$1/x^2$ interaction, can be viewed as a coordinate description of these
generalized statistics. Some of the peculiarities of this system are briefly
discussed and the question is addressed whether there also for the
$N$-particle system is some underlying fundamental algebra, such that different
values of the statistics parameter correspond to inequivalent representations
of this algebra. Such a parameter independent algebra has indeed been shown to
be present in the Calogero model \cite{Isakov 96}. I will discuss how this
infinite-dimensional algebra can be constructed by use of operators from the
Calogero model and show that the algebra can be defined in a more abstract way.
This re-formulation of the algebra makes it possible to identify the presence
of the same algebra in another, but related system, the matrix model. In this
model the variables are $N\times N$ Hermitian matrices and the model can be
related to the Calogero model by fixing some of the constants of motion. I
conclude with some comments about unsolved problems and possible physical
applications of the algebra.

The Heisenberg quantization scheme for identical particles, which is discussed
here, has been developed together with Jan Myrheim \cite{Leinaas 88,Leinaas
91}. The infinite-dimensional algebra of the Calogero model has been studied in
a paper with Serguei Isakov \cite{Isakov 96}, and finally the simplifications
in the formulation of the algebra and the application to the matrix models have
been discussed in a work with Jan Myrheim, Serguei Isakov, Alexios
Polychronakos and Raimund Varnhagen \cite{Isakov 96b}. 

\section{The algebraic Heisenberg quantization scheme}
In Heisenberg's formulation of quantum mechanics the commutation relation
between position and momentum is the basic ingredient \cite{Heisenberg 25}. It
{\em defines} the quantum mechanics of the system in the sense that it can be
used to derive the matrix elements of any observable, \ie any function of
position and momentum. Heisenberg's canonical commutation relation can be
viewed as defining a particular Lie algebra. This Lie algebra is represented in
the classical description of the system in the form of Poisson brackets as well
as in the quantum description in the form of commutators. The quantum
mechanical system can be seen as defining a represention of this fundamental
{\em Lie algebra of observables} in terms of a commutation algebra of Hermitian
operators in a Hilbert space.

Viewed in this way we may formulate the Heisenberg approach to quantum
mechanics as a general algebraic program for the quantization a classical
system.  The program has the following ingredients: \\ 

\noindent
{\em a)} Choose a complete set of fundamental observables $ A,B,...$ which form
a closed Lie algebra under Poisson brackets. \\ 
{\em b)} Represent the
observables as Hermitian operators $ \hat A, \hat B,...$ which define an
irreducible representation of the algebra, with Poisson brackets replaced by
commutators, $\{A,B\}_{pb}\rightarrow \frac{1}{i\hbar}[\hat A,\hat B]$. \\ 
{\em c)} Represent a general observable as a function of the fundamental
observables
$ \hat A, \hat B,...$. \\

Since the representation is irreducible (which reflects the fact that the set of
observables is complete), the eigenvalues and matrix elements (in a chosen
basis) of all the observables are determined by their algebraic relations. In
the case of the (Heisenberg-Weyl) algebra of position and momentum, the
irreducible representation is unique. More generally we expect that several,
inequivalent representations may exist. These different representations can be
interpreted as inequivalent quantizations of the same classical system. As I
will discuss, the different particle statistics appear in this approach as
inequivalent representations of the same algebra of observables for a system of
identical particles.

Clearly the program for quantization referred to above is somewhat loosely
defined. There are several remaining questions and possible ambiguities. One
possible ambiguity has to do with the choice of a fundamental set of
observables. Different choices may lead to different results. Another ambiguity
has to do with the operator ordering problem. This makes the mapping from the
classical to the quantum observables non-unique. It may also affect the
fundamental algebra itself in the form of quantum corrections to the
commutation relations. And finally there may be a conflict between the number
of fundamental observables which is needed to close the Lie algebra and the
number of independent variables of the (classical) system.

It is not my intention to make any attempt to give a more precise formulation
of the general program. After all we have to accept that there are inherent
ambiguities in the quantization of a classical system. However, for
sufficiently simple systems the problems and ambiguities may be less important.
I will first demonstrate this for two simple cases of a single particle, and
then turn to the question of what happens for a system of identical particles
in one dimension.

\section{Systems of non-identical particles} For a system of non-identical
particles there is an independent set of observables for each particle of the
system. This means that the $N$-particle (Hilbert) space is a product space of
single particle spaces. The quantization of the system therefore can be
acomplished by quantizing the single particle systems separately. Interactions
may certainly make the dynamics of the $N$-particle system complicated, but
this we do not see as important for the question of how to quantize the system.
(Note, however, that this view may be too simpleminded if there are singular
interactions or velocity-dependent forces present.) With this argument in mind,
I will briefly examine two elementary systems where a single particle moves in
one dimension, first on an infinite line and then on a circle.

\subsection{One particle on a line -- the Heisenberg-Weyl algebra}
This first example is almost too simple. It is well-known from introductory
quantum mechanics  and it is the original case discussed by Heisenberg, Born and
Jordan \cite{Heisenberg 25} in the form of a fundamental commutation relation
between position and momentum. However, since this is the model for what I will
also later discuss, I will nevertheless include a brief discussion.

The natural set of fundamental observables in this case consists of the
position $x$ and the momentum $p$. If we extend this by a third observables,
which I will denote $\lambda$, we have a closed Lie algebra,
\be
[x,p]=i\lambda, \nonumber  \\\ 
[x,\lambda]=[p,\lambda]=0
\ee
In an irreducible representation $\lambda$ is proportional to the identity, an
the standard form of the canonical quantization condition is obtained if we
write it as
\be
\lambda=\hbar \, {\bf 1}
\ee
(Note that other values of $\lambda$ do not correspond to different
representatitons, since they can be related by a rescaling of $x$ and $p$.)

An irreducible representation of the algebra is most readily constructed in the
standard way by use of the raising and lowering operators 
\be
a=\frac{1}{\sqrt{2\mu}}\left(x+\frac{i}{\hbar}\mu p\right) \nonumberÊ\\
a^\dagger=\frac{1}{\sqrt{2\mu}}\left(x-\frac{i}{\hbar}\mu p\right)
\ee
where $\mu$ is an arbitrary parameter of the correct dimension.
Hermiticity requires a lowest state annihilated by $a$
\be
a \left| 0 \right\rangle =0
\ee
and a complete set of basis states is constructed iteratively by use of
$a^\dagger$,
\be
a^\dagger\left|n\right\rangle=\sqrt{n+1}\left|n+1\right\rangle
\ee
The construction shows that the irreducible representation is unique.

The connection to the Schr\"odinger formulation is obtained if we introduce a
orthonormalized set of position states, the eigenstates of $x$. In this
coordinate representation $p$ is represented, in the usual way, as the
differential operator
$\frac{\hbar}{i}\frac{d}{dx}$.

As already mentioned the generalization to several particles, and also to higher
dimensions, is straight forward, since there is an independent algebra for each
particle and each dimension.

\subsection{One particle on a circle}
If the particle moves on a circle rather than on a straight line, it is well
known from the Schr\"odinger approach that a new feature appears. The wave
functions will pick up a complex phase factor when the circle is traversed.
This phase factor is not specified by the classical theory, and different
values correspond to inequivalent quantizations of the system. A physical
interpretation is that the particle is charged and that the complex phase is
related to a magnetic flux which penetrates the circle.

Let us examine the system from the algebraic point of view. We first note that
the polar angle should not be considered as an observable for the position of
the particle. This is because it is not invariant when the particle completes a
full cycle. Let us instead use the cartesian coordinates $x$ and $y$ as
observables for the position. Together with the angular momentum $L$ they form
a closed Lie algebra,
\be
[L,x]=i\hbar\, y \;\;\;\; [L,y]=-i\hbar\, x \;\;\;\; [x,y]=0
\ee
Apparently we have too many observables to describe the two-dimensional phase
space, but there is an invariant, which in an irreducible representation takes
a constant value,
\be
x^2+y^2=r^2 {\bf 1}
\ee
The constant $r$ is naturally interpreted as the radius of the circle and the
relation reduces the number of independent variables to two.

Also in this case the irreducible representations are most conveniently
constructed by use of raising and lowering operators, but now we choose
$x_{\pm}=x\pm iy$. They satisfy
\be
[L,x_{\pm}]=\pm \hbar x_{\pm}
\ee
and therefore are climbing operators in the spectrum of $L$. Beginning from an
arbitrary eigenstate of $L$, a tower of eigenstates is constructed, which span
the Hilbert space,
\be
L \left| m \right\rangle =(m+\alpha) \hbar \left| m \right\rangle \nonumber \\ 
x_\pm \left| m \right\rangle = r \left| m \pm 1 \right\rangle 
\ee
The parameter $m$ takes all integer values, and the spectrum of $L$ is unbounded
both from above and below. The parameter $\alpha$, which describes a shift of
the spectrum relative to the integer values is undetermined by the algebraic
relations. It characterizes inequivalent representations of the algebra.

A coordinate representation can be introduced in terms of the eigenstates of the
position operators $x_\pm$ and the corresponding wave functions,
\be
x_\pm\left|\phi\right\rangle=re^{\pm i\phi}\left|\phi\right\rangle \nonumber \\
\psi(\phi)=\langle\phi|\psi\rangle
\ee
The form of the angular momentum operator in the coordinate representation can
be deduced from the commutation relations. If we choose the position
eigenvectors and therefore the wave functions to be single valued on the
circle, the operator is found to be
\be
L=-i\hbar \left(\frac{d}{d\phi}+\alpha\right)
\ee
Clearly this result agrees with the Schr\"odinger quantization of the system.
The parameter $\alpha$ in that case appears as through the periodicity
condition on the wave functions. In the present case the parameter defines
inequivalent irreducible representations of the same fundamental algebra of
observables.

\section{Identical particles and the $1/x^2$ interaction} 
For identical particles there is a restriction on the observables relative to
the case of non-identical particles. The observables should be symmetric with
respect to particle indices. We consider first the implication of this for the
case of two identical particles on the (infinite) line \cite{Leinaas 88}. The
center-of-mass coordinate $X=\half(x_1+x_2)$ and momentum
$P=p_1+p_2$ are not affected by the symmetry constraint, and they define the
same Heisenberg-Weyl algebra as that of a single particle. However, the
constraint is important for relative coordinate and momentum. The linear
variables $x=x_1-x_2$ and
$p=\half(p_1-p_2)$ are not observables since they are not symmetric, but the
quadratic variables are, and they form a closed Lie algebra. If we choose the
following combinations
\be
A=\frac{1}{4}(p^2+x^2) \;\;\; B=\frac{1}{4}(p^2-x^2) \;\;\;
C=\frac{1}{4}(px+xp) 
\ee 
and introduce the Hermitian conjugate operators
\be
B_\pm=B\pm iC
\ee
these operators satisfy the standard commutation relations of the $su(1,1)$ or
$sp(2,R)$ algebra
\be
[A,B_\pm]=\pm B_\pm \;\;\;\; [B_+,B_-]=-2A
\ee
(In these expressions I have assumed dimensionless variables $x$ and $p$.) 

The quadratic variables $A$, $B$ and $C$ (together with the CM variables $X$ and
$P$) may be considered as the fundamental observables of the system. Following
the general approach outlined above we now quantize the system by constructing
irreducible representations of the fundamental algebra of observables. (In the
following we suppress the CM coordinate and momentum since they can easily be
handled as in the single particle case.)

The representations can be constructed in the same way as for the particle on
the circle. We construct a tower of eigenstates of the operator $A$ by use of
the lowering and raising operators $B_-$ and $B_+$. Hermiticity in this case
implies that there is a lowest state annihilated by $B_-$, and for general eigen
vectors of $A$ we find the relations
\be
B_- \left|n\right\rangle &=&\sqrt{n(n-1+2\alpha)}\left|n-1\right\rangle
\nonumber \\
B_+\left|n\right\rangle&=&\sqrt{(n+1)(n+2\alpha)}\left|n+1\right\rangle
\nonumber \\ A\left|n\right\rangle&=&(\alpha+n)\left|n\right\rangle
\ee
These relations define an infinite basis for the representation and specify the
matrix elements of the fundamental observables. We note that again a (real)
parameter
$\alpha$  appears, which characterizes inequivalent representations of the
algebra. With $n$ restricted to non-negative integers, the parameter $\alpha$
is restricted (by Hermiticity of the observables) to be positive. Different
values of $\alpha$ we interpret as corresponding to inequivalent quantizations
of the two-particle system.

The spectrum of $A$, which is identical to a harmonic oscillator spectrum,
shows that the parameter value $\alpha=1/4$ corresponds to a system of bosons,
while $\alpha=3/4$ corresponds to fermions. One may note that there is no exact
periodicity in $\alpha$, but there is a quasi-periodicity in the sense that all
levels except the lowest one are reproduced when $\alpha$ increases by 1.

Also in this case a coordinate representation may be introduced in terms of
eigenvectors of the operator $x^2$. From the action of the operators on these
states, one finds expressions for the observables $A$, $B$ and $C$ in terms of
differential operators which are unusual in the sense that the operator $p^2$
includes a singular potential,
\be
p^2=-\frac{d^2}{dx^2}+\frac{\lambda}{x^2} ,
\ee
where strength of the potential $\lambda$ is determined by the parameter
$\alpha$,
\be
\lambda=4\left(\alpha - \frac{1}{4}\right)\left(\alpha - \frac{3}{4}\right)
\ee
This means that the kinetic energy of the two-particle system in the general
case will include a ``statistical" $1/x^2$-potential. This potential vanishes
only for bosons and fermions.

Before proceeding, I would like to include some comments about the $1/x^2$
potential. There are several features of a system of particles with this
interaction which are similar to those of a free particle system (see
ref.\cite{Polychronakos 89} for a detailed discussion). Let us first consider a
two-particle collision. A characteristic feature of this collision is that
there is no time delay. This means that the particles after the collision
(asymptotically) move as if they just passed through each others. Only at close
distance the particle trajectories deviate from those of non-interacting
particles. In the quantum description of the collision one finds that there is
a constant phase shift
\be
\phi_{sc}=-2 \pi \left(\alpha-\frac{1}{4}\right)
\label{phase}
\ee
The momentum independence of this phase shift is directly related to the lack of
time delay in the collision.

These features in fact carry over to the $N$-particle system. The effect of
scattering of the classical system is just to permute the momenta of the
incoming particles in the out state, and there is no time delay relative to the
motion of non-interacting particles. In the quantum case the overall phase
shift is again momentum independent and it is simply the sum of two-particle
contributions (\ref{phase}) from each pair of particles in the system. 

Let us next consider what happens when a harmonic oscillator interaction is
added to the kinetic term. For two particles we already know the result, since
the Hamiltonian is essentially the observable $A$ of the $su(1,1)$ algebra. The
effect of the $1/x^2$-interaction is simply to shift all the energy levels so
that the constant separation between the levels is preserved. For the
N-particle system the Hamiltonian gets the form
\be
H=\half\sum_{i=1}^N\left(-\frac{\partial^2}{\partial x_i^2}+2\sum_{i\neq j}
\frac{\lambda}{\left(x_i-x_j\right)^2}+x_i^2\right)
\ee 
This interacting model is known as the Calogero model and it has been
extensively studied since it was first presented and shown to be exactly
solvable \cite{Calogero 69}. Again the effect of the interaction is to give an
overall shift of the whole energy spectrum, so that the energies can be written
in the form
\be
E=E_B + \left(\alpha - \frac{1}{4}\right) N(N-1)
\ee
where $E_B$ is the corresponding energy of the bosonic $N$-particle harmonic
oscillator. (The energy is here scaled with the harmonic osciallator frequency
to be a dimensionless quantity.)

\section{An algebra for more than 2 particles}
So far I have restricted the discussion of the fundamental algebra of
observables to the case of two identical particles in one dimension. The
generalization to higher dimensions (but still only two particles) is straight
forward in the sense that it is easy to identify the algebra. The quadratic
variables in the relative coordinate and momentum define the Lie algebra
$sp(2d,R)$, where $d$ is the dimension of the one-particle configuration space.
These variables generate the linear symplectic transformation in the
$2d$-dimensional phase space. Representations of these algebras may be
constructed in a similar way as discussed here for $d=1$, but the construction
becomes more complicated with additional degeneracies of the energy levels.
With Jan Myrheim I have examined the case of two particles in two dimensions in
some detail. Also in this case we find representations different from those
which describe bosons and fermions, but the interpretation is not as simple as
in one dimension, since additional degrees of freedom appear in the general
case. I will only refer to \cite{Leinaas 91} for a discussion of the
two-dimensional case, while here returning to the question of the algebra of
observables for a system of identical particles in one dimension
\footnote{Note that in \cite{Leinaas 91} the notation $sp(d,R)$ is used for the
algebra instead of $sp(2d,R)$.}.

When the number of particles is larger than $2$ the symmetric polynomials in
$x$ and $p$ of degree lower or equal to $2$ do not form a complete set of
observables. There are symmetric functions of $x$ and $p$ that cannot be
expressed in terms of these variables alone. Higher order polynomials are
needed, but may be restricted to a degree lower or equal to $N$. However, the
problem then arises that these polynomials do not form a closed
finite-dimensional Lie algebra. We have the choice either to consider an
algebra which is non-linear (\ie products of the generators will appear as the
result of commutation), or we extend the algebra to an infinite-dimensional
algebra by including polynomials of arbitrarily high degree while neglecting all
algebraic relations between the polynomials that exist for a finite particle
number $N$. The latter point of view is only the one which sems possible to
handle and it is this view I will adopt in the following. More specifically the
algebra is assumed to be spanned by polynomials of the one-particle form
\be
L_{mn}=\sum_{i=1}^N x_i^m p_i^n
\ee
where in the quantum case the ordering of the operators becomes important. This
set of variables is large enough to generate all (symmetric) observables in $x$
and $p$. The philosophy is that the algebraic relations between these variables
, which appear for finite $N$ are not included in the definition of the
algebra, but rather appear at the level of irreducible representations. In
particular the value of the particle number will be specified at this level,
since the algebra is independent of the number of particles in the system.

The classical Lie algebra is defined by the Poisson brackets, which have the
form
\be
\{L_{kl},L_{mn}\}_{pb}=(kn-ml)L_{k+m-1,l+n-1}
\ee
We may interpret the variables $L_{mn}$ as generators for canonical
transformations in a two-dimensional phase space.  The infinite-dimensional Lie
algebra in fact generate the full group of (continuous) canonical
transformations. 

The canonical transformations are area-preserving, and for a system of
particles this means that the transformations preserve the particle density.
This has been noted as interesting for the description of the incompressible
states of the quantum Hall system, where the electrons have effectively a
two-dimensional phase space due to the strong magnetic field. In a classical
description of this incompressible fluid the infinite-dimensional algebra has
the effect of generating perturbations of the edge of the system \cite{Cappelli
93}. The classical Lie algebra defined by the generators $L_{mn}$ is sometimes
referred to as $w_\infty$. (More precisely it may be referred to as the positive
part of the algebra, which then is defined to include generators $L_{mn}$ also
for negative $m,n$.) 

A precise definition of the quantum observables $L_{mn}$ depends on a
specification of the operator ordering of $x$ and $p$ in the product. However, a
reordering gives an operator which can be expressed as a linear combination of
operators whith the original ordering. This means that we can pick an arbitrary
ordering of operators for each $m,n$ and view a reordering as a change of basis
of the algebra which is defined by these operators.

The Lie algebra of the quantum observables are defined by the commutation
relations between the operators $L_{mn}$. For bosons as well as for fermions
they have the form
\be
\left[L_{kl},L_{mn}\right]&=&i\hbar(km-ln)L_{k+m-1,l+n-1} \nonumber \\
   &+&i\hbar^3c_{klmn} L_{k+m-3,l+n-3} \nonumber \\
&+& \;\;\;...
\ee
We note that there are quantum corrections (higher order in $\hbar$) relative to
the classical algebra. The precise form of these correction terms depend on the
choice of operator ordering, but the point is that these new terms cannot all be
absorbed by a redefinition of the operators $L_{mn}$. The algebra of the quantum
system, which is often referred to as (the positive part of) $W_{1+\infty}$ is
then different from the classical algebra $w_\infty$.

I would now like to consider the question of generalizations of bosons and
fermions in a similar way as already discussed for the 2-particle case. The
fundamental algebra of observables then plays a central role, and the main
point I will discuss is the question of how to identify this algebra. The
generators will be assumed to be of the form $L_{mn}$. Clearly the classical
algebra cannot be identical to this commutation algebra, since it does not even
reproduce bosons and fermions. Another possibility would be to assume the
algebra $W_{1+\infty}$ of the boson and fermion system to be the fundamental
algebra and then to look for other representations to describe systems of
particles with generalized statistics. Such generalized representations have
already been considered in the context of the quantum Hall system
\cite{Cappelli 94}. However, a different view is adopted in \cite{Isakov 96}
and will be discussed here. As already pointed out for the case of two
particles, the generalized statistics found in the algebraic approach can be
interpreted as a (statistical) $1/x^2$ interaction between the particles. This
description of the two-particle system is naturally extended to a system of $N$
particles and suggests that the Calogero model can be viewed as describing a
system of identical particles with generalized (one-dimensional) statistics.
Based on this view we make the assumption that the fundamental algebra is
already represented in the Calogero model. This model can then be used to
identify and to construct the algebra.

\section{A parameter independent algebra for the Calogero model}

As already noted for the 2-particle case there exist representations of the
symmetrized quadratic polynomials in $x$ and $p$ that do not admit these
observables to be expressed as functions of the single-particle (and
non-observable) variables $x_i$ and $p_i$, where $i$ is the particle index.
However, as shown in \cite{Polychronakos 92,Brink 92} this is true only if we
insist that $x_i$ and $p_j$ should satisfy the standard Heisenberg commutation
relations. A special modification of this algebra may be introduced which
depends on the  (statistics) interaction  parameter and  where the observables
are expressed as the standard symmetric sum over products of single-particle
operators. Remarkably this algebra can be extended to an algebra for the full
$N$-particle system and it defines a spectrum generating algebra for the
Calogero Hamiltonian.

The modified Heisenberg algebra has been referred to as an $S_N$-extended
Heisenberg algebra, since it includes permutation operators in addition to $x$
and
$p$. The form of the modified commutator is (with $x$ and $p$ again in
dimensionless units) 
\be
\left[x_i,p_j\right]=i(\delta_{ij}(1+\nu\sum_{k\neq i}K_{ik})-\nu
K_{ij}) 
\label{modcom}\ee
where $K_{ij}$ are transposition operators that interchanges the particle
indices
$i$ and $j$,
\be
K_{ij}x_j=x_iK_{ij} \;\;\;\; K_{ij}p_j=p_i K_{ij}
\ee
and further satisfy a set of product relations characteristic for the
permutation group. 

The parameter $\nu$ which is present in the modified commutator (\ref{modcom})
is related to the statistics parameter $\alpha$ which was introduced at an
earlier stage,
$\nu=2\alpha - 1/2$. The strength of the statistical interaction in this new
parameter is 
\be
\lambda=4(\alpha-\frac{1}{4}))(\alpha-\frac{3}{4}))=\nu(\nu-1)
\ee
In the following I will use the parameter $\nu$ instead of $\alpha$ or
$\lambda$.

To stay close to the notation of ref.\cite{Isakov 96} I will now change to the
(Hermitian conjugate) variables 
\be
a_i=\frac{1}{\sqrt{2}}(x_i+ip_i)
\;\;\;\;a_i^\dagger=\frac{1}{\sqrt{2}}(x_i-ip_i)
\ee 
instead of $x_i$ and $p_i$ and introduce the observables $L_{mn}$ in a slightly
changed way as
\be
L_{mn}^\alpha\approx\sum_{i=1}^Na_i^{\dagger m}a_i^n
\ee
In this equation $\approx$ means equal up to operator ordering. We use the
additional index $\alpha$ to distinguish between operators 
which differ by orderings of the operators $a_i$ and $a_i^\dagger$. It is the
algebra of the observables $L_{mn}^\alpha$ which is the object of our interest,
and by use of the (parameter-{\em dependent}) $S_N$-extended Heisenberg algebra I
will sketch how a parameter-{\em independent} algebra of observables can be
constructed. For further details I refer to \cite{Isakov 96}.

Before we examine further this algebra of observables it is of interest to
note that the Calogero Hamiltonian belongs to this algebra and is
essentially identical to the operator $L_{11}$. (Actually, this is after a
factor $\prod (x_i-x_j)^\nu$ has been separated out of the wave functions.)
This operator has simple commutators with the operators
$a_i$ and
$a_i^\dagger$,
\be
[H,a_i^\dagger]=a_i^\dagger
\label{Hcom}
\ee
This implies that the eigenstates and eigenvalues of $H$ can be constructed in a
similar way as for the ordinary harmonic oscillator. Note however that when
applying the non-symmetrized operators $a_i^\dagger$ to build up the space of
states one constructs an extension of the original Calogero model. The original
model is defined on the subspace of symmetrized or of anti-symmetrized states.
(For the same reason this extended model will define a reducible representation
of the algebra of observables $L_{mn}^\alpha$).

To define the algebra of observables I will follow the construction of
ref.\cite{Isakov 96}. In this approach one defines rules for constructing the
generators, and by repeatedly applying the rules an infinite-dimensional basis
of generators is built up. Since the rules can be shown to generate
$\nu$-independent operators, the corresponding algebra will be parameter
independent. Unfortunately this does not mean that the algebra can be expressed
in a closed form, but rather that we have a method to construct a basis for the
algebra step by step.

The basic building blocks of the construction are the variables 
\be
L_{m0}=\sum_ia_i^{\dagger m} \;\;\;\; L_{0n}=\sum_ia_i^n 
\ee
These operators are assumed to belong to the algebra and they are not affected
by the operator ordering problem. (More precisely it is the Hermitian
combinations of the operators that are elements of the algebra.) By commuting
these operators we may now generate new elements of the algebra. If
$A_k$ denote an operator of the above type, we will after $n$ steps produce a
string of commutators of the form
\be
S=[A_n, \;...,[A_3[A_2,A_1]]\;...]
\ee
(In the following I will refer to such a repeated commutator simply as a
``string".) The commutator of two strings can be expressed as a sum of other
strings by use of the Jacobi identity. If therefore all strings are included,
with the number $n$ of operators arbitrarily large, the construction clearly
defines an infinite-dimensional Lie algebra. This algebra is the one we seek.

There is an element we have to specify in this construction. If we calculate the
commutators by repeated use of the fundamental commutation relation
(\ref{modcom}) we can bring any string into a standard form (\eg normal ordered
form). If all these expressions should be independent of $\nu$ (and $K_{ij}$),
the resulting algebra would be identical to the algebra $W_{1+\infty}$ of
bosons and fermions. However, by examining a few examples we can convince
ourselves that all $\nu$-dependent terms cannot be avoided if we introduce a
unique operator ordering for a given product of $a_i$ and $a_i^\dagger$. This
means that the algebra $W_{1+\infty}$ is not represented in the Calogero model
(at least not in this way).

To be able to proceed  we have to make the basic assumption that operators
$L_{mn}^\alpha$ that differ by reordering of operators $a_i$ and $a_i^\dagger$
should be considered as independent elements of the algebra. This means that
when defining the (abstract) algebra we simply neglect  $\nu$-dependent
relations that are present in the Calogero model. Again we make the assumption
that such relations appear only at the level of irreducible representations.
Some identitities between reordered operators nevertheless have to survive to
the level of the algebra. These are the ones needed to satisfy the Jacobi
identity. Our precise definition of the algebra is that all $\nu$-{\em
independent} identities which are present in the Calogero model will be imposed
on the elements of the algebra. Even if it is clear that an
infinite-dimensional algebra can be constructed in this way, it is still a
non-trivial question whether this algebra will be independent of the statistics
parameter $\nu$. However, that is in fact the case, as has been demonstrated in
\cite{Isakov 96}, and I will now outline how this can be shown. 


Let us consider the commutator between an arbitrary observable of the form
$L_{kl}^\alpha$ and an observable $L_{0n}$. It can be written in the form
\be
[L_{0n},L_{kl}^\alpha]=\sum_i\sum_j\sum_p {\cal A}_j^p\;[a_i^n,a_j^\dagger]\;
{\cal B}_j^p
\label{Lcom}
\ee
where${\cal A}_j^p$ and ${\cal B}_j^p$ denote products of operators $a_j$ and
$a_j^\dagger$ only referring to particle $j$. The commutator appearing in the
expression on the right-hand side can be evaluated
\be [a_i^n, a_j^{\dagger}]=n
\delta_{ij} a_i^{n-1} + \nu
\delta_{ij}\sum_l\sum_{s=0}^{n-1}a_i^{n-1-s}a_l^s K_{il}
-\nu\sum_{s=0}^{n-1}a_i^{n-1-s}a_j^s K_{ij} \,. \label{} 
\ee
When this is inserted in the equation above, the $\nu$-dependent terms vanish
due to the summation over $i$ and $j$. This important result we write as
\be
\sum_{ij} {\cal A}_j\; [a_i^r,a_j^{\dagger}]\;{\cal B}_j=\sum_i {\cal A}_i\;
ra_i^{r-1}\;{\cal B}_i \,,
\label{7}
\ee
This is the basic relation needed to show the $\nu$-independence of the algebra.
We note that the right-hand-side is a new one-particle operator. This implies
that the commutator (\ref{Lcom}) can be written in the form
\be
[L_{0n},L_{kl}^\alpha]=\sum_\beta
c^\alpha_\beta(n,kl)L_{k-1,n+l-1}^\beta
\label{LL}
\ee
where the coefficients $c^\alpha_\beta(n,kl)$ are $\nu$-independent. All
$\nu$-dependence has been absorbed in the different orderings of operators in
the observables $L_{k-1,n+l-1}^\beta$. 

Clearly a commutator which involves $L_{n0}$ instead of $L_{0n}$ can
be written in a similar form. From this we conclude that a string of
commutators of these two types of operators can be reduced to the form
(\ref{LL}). This means that any string can be expressed as a linear
combination of operators
$L_{kl}^\alpha$ with $\nu$-independent coefficients. Since the algebra has been
defined to be spanned by this set of strings, this implies that algebra is
independent of the statistics parameter $\nu$.

\section{A step by step construction}
The situation then is that we know that a parameter independent algebra exists
and that it can be constructed by repeated commutation of the basic elements
$L_{m0}$ and $L_{0n}$. This does not mean that we know the full structure of
the algebra, but we are able to construct it in a stepwise fashion. To do the
commutations is straight forward, but to check for identities between operators
which correspond to reordered products of $a_i$ and $a_i^\dagger$ is
non-trivial, since there is no obvious simple and systematic method to do this.
The low-order part of the algebra has nevertheless been studied, in
\cite{Isakov 96} by use of the operators of the Calogero model, and in
\cite{Isakov 96b} by use of the matrix  representation which I will discuss
later.

To classify the generators it is convenient to introduce the notion of a spin
index. The three operators $\half L_{01}, \half L_{11}$ and
$\half L_{10}$ define an $su(1,1)$ subalgebra. It acts in the adjoint
representation (\ie by commutation) on the other generators of the algebra, and
the full algebra can be divided into finite dimensional representations of the
subalgebra. This means that the generators can be characterized by a spin
quantum number. The raising and lowering operators $\half L_{10}$ and $\half
L_{01}$ mix operators $L_{mn}$ with the same value of $m+n$, while the
commutator with
$\half L_{11}$ determines the ``z-component" of the spin,
\be
[\half L_{11},L_{mn}]=\half (m-n)L_{mn}
\ee
By counting the number of states, one readily sees that
$\half (m+n)$ is the maximum spin associated with an operator $L_{mn}$. For the
lowest values of
$m+n$ the maximum spin multiplet is the only multiplet present, but for higher
values of $m+n$ there are an increasing number of multiplets of lower spin. The
number of multiplets can be determined from the list of degeneracies, \ie from
the number of independent operators $L_{mn}$ for given $(m,n)$. In fig.1 a list
of degeneracies is given for the lowest values of $m+n$. 


\begin{figure}[htb]
\unitlength=0.75mm
\thicklines
\begin{center}
\begin{picture}(120,160)
\put(10,10){\vector(1,0){145}}
\put(10,10){\vector(0,1){145}}
\put(20,9){\line(0,1){2}}
\put(35,9){\line(0,1){2}}
\put(50,9){\line(0,1){2}}
\put(65,9){\line(0,1){2}}
\put(80,9){\line(0,1){2}}
\put(95,9){\line(0,1){2}}
\put(110,9){\line(0,1){2}}
\put(125,9){\line(0,1){2}}
\put(140,9){\line(0,1){2}}
\put(9,20){\line(1,0){2}}
\put(9,35){\line(1,0){2}}
\put(9,50){\line(1,0){2}}
\put(9,65){\line(1,0){2}}
\put(9,80){\line(1,0){2}}
\put(9,95){\line(1,0){2}}
\put(9,110){\line(1,0){2}}
\put(9,125){\line(1,0){2}}
\put(9,140){\line(1,0){2}}
\put(160,10){\makebox(0,0){$n$}}
\put(10,160){\makebox(0,0){$m$}}
\put(20,20){\begin{picture}(100,100)
\put(0,0){\makebox(0,0){$1$}}
\put(0,-15){\makebox(0,0){$0$}}
\put(15,-15){\makebox(0,0){$1$}}
\put(30,-15){\makebox(0,0){$2$}}
\put(45,-15){\makebox(0,0){$3$}}
\put(60,-15){\makebox(0,0){$4$}}
\put(75,-15){\makebox(0,0){$5$}}
\put(90,-15){\makebox(0,0){$6$}}
\put(105,-15){\makebox(0,0){$7$}}
\put(120,-15){\makebox(0,0){$8$}}
\put(-15,0){\makebox(0,0){$0$}}
\put(-15,15){\makebox(0,0){$1$}}
\put(-15,30){\makebox(0,0){$2$}}
\put(-15,45){\makebox(0,0){$3$}}
\put(-15,60){\makebox(0,0){$4$}}
\put(-15,75){\makebox(0,0){$5$}}
\put(-15,90){\makebox(0,0){$6$}}
\put(-15,105){\makebox(0,0){$7$}}
\put(-15,120){\makebox(0,0){$8$}}
\put(15,0){\makebox(0,0){$1$}}
\put(30,0){\makebox(0,0){$1$}}
\put(45,0){\makebox(0,0){$1$}}
\put(60,0){\makebox(0,0){$1$}}
\put(75,0){\makebox(0,0){$1$}}
\put(90,0){\makebox(0,0){$1$}}
\put(105,0){\makebox(0,0){$1$}}
\put(120,0){\makebox(0,0){$1$}}
\put(0,15){\makebox(0,0){$1$}}
\put(15,15){\makebox(0,0){$1$}}
\put(30,15){\makebox(0,0){$1$}}
\put(45,15){\makebox(0,0){$1$}}
\put(60,15){\makebox(0,0){$1$}}
\put(75,15){\makebox(0,0){$1$}}
\put(90,15){\makebox(0,0){$1$}}
\put(105,15){\makebox(0,0){$1$}}
\put(0,30){\makebox(0,0){$1$}}
\put(15,30){\makebox(0,0){$1$}}
\put(30,30){\makebox(0,0){$2$}}
\put(45,30){\makebox(0,0){$2$}}
\put(60,30){\makebox(0,0){$3$}}
\put(75,30){\makebox(0,0){$3$}}
\put(90,30){\makebox(0,0){$4$}}
\put(0,45){\makebox(0,0){$1$}}
\put(15,45){\makebox(0,0){$1$}}
\put(30,45){\makebox(0,0){$2$}}
\put(45,45){\makebox(0,0){$3$}}
\put(60,45){\makebox(0,0){$4$}}
\put(75,45){\makebox(0,0){$5$}}
\put(0,60){\makebox(0,0){$1$}}
\put(15,60){\makebox(0,0){$1$}}
\put(30,60){\makebox(0,0){$3$}}
\put(45,60){\makebox(0,0){$4$}}
\put(60,60){\makebox(0,0){$7$}}
\put(0,75){\makebox(0,0){$1$}}
\put(15,75){\makebox(0,0){$1$}}
\put(30,75){\makebox(0,0){$3$}}
\put(45,75){\makebox(0,0){$5$}}
\put(0,90){\makebox(0,0){$1$}}
\put(15,90){\makebox(0,0){$1$}}
\put(30,90){\makebox(0,0){$4$}}
\put(0,105){\makebox(0,0){$1$}}
\put(15,105){\makebox(0,0){$1$}}
\put(0,120){\makebox(0,0){$1$}}
\end{picture}}

\end{picture}
\end{center}
\caption{The degeneracies of operators $L_{mn}$ for the points $(m,n)$ with
$m+n\leq 8$.}
\label{fig1}
\end{figure}
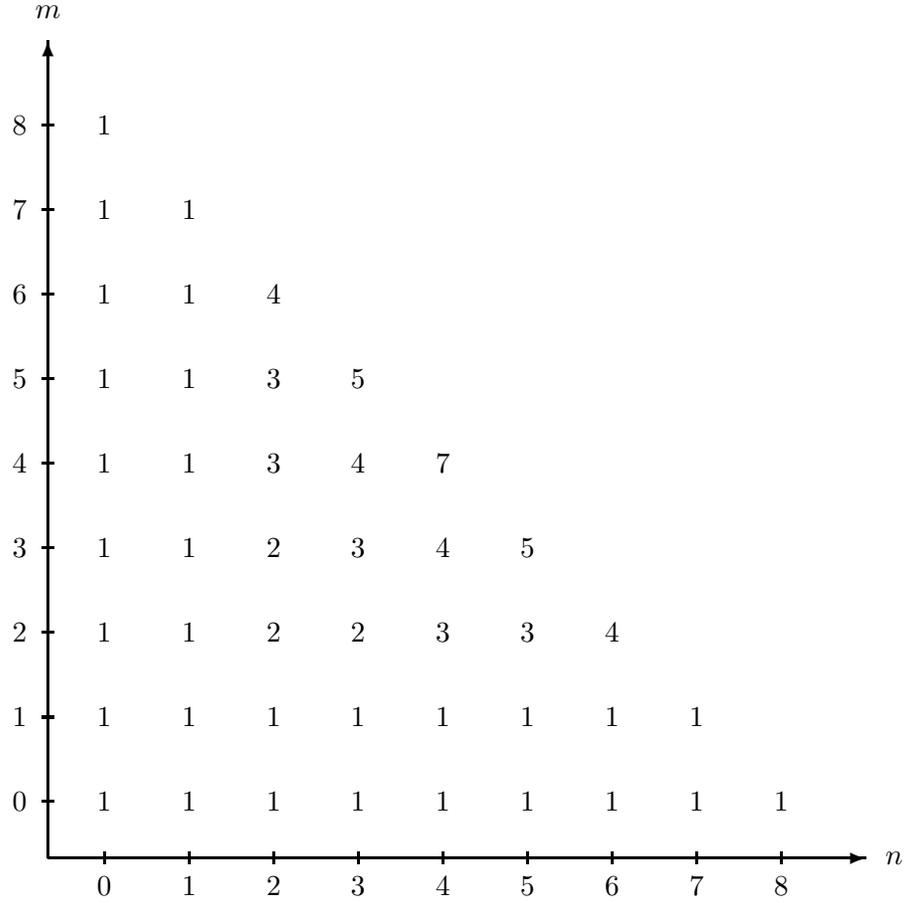


I will here give some of the expressions for the generators for the lowest
values of $m+n$. Since they all are operators of the one-particle form I may for
simplicity suppress the particle index and the summation over this index. It is
also sufficient to list the generators only for $m\geq n$ due to the Hermiticity
relation $L_{nm}=L_{mn}^\dagger$. For $m+n\leq 3$ only the maximal spin
multiplets exist, with values running from $0$ to $3/2$. The expressions are
\be 
s=0: \;\;\;\; L_{00}&=&1 \nonumber \\ s=\half: \;\;\;\; L_{10}&=&a^\dagger
\nonumber \\ s=1: \;\;\;\; L_{20}&=&a^{\dagger 2}\;\;\;\;L_{11}=\half(a^\dagger
a+a a^\dagger)
\nonumber \\  s=\frac{3}{2}: \;\;\;\; L_{30}&=&a^{\dagger
3}\;\;\;\;L_{21}=\half(a^{\dagger 2} a+a a^{\dagger 2})
\ee
For $m+n=4$ there are two multiplets, of spin $2$ and $0$. The spin $2$
operators are
\be
L_{40}^2&=&a^{\dagger 4}\;\;\;\;L_{31}^2=\half(a^{\dagger 3} a+a a^{\dagger 3})
\;\;\;\;L_{22}^2=\frac{1}{6}(a^{\dagger 2} a^2+a^\dagger a a^\dagger a+a^\dagger
a^2 a^\dagger + sym) 
\ee 
where $sym$ means the symmetric expression by right-left reflection of the
products. The spin $0$ operator is
\be
L_{22}^0=L_{31}=a^{\dagger 2} a^2+a^\dagger a a^\dagger a + sym  
\ee
The way these and other operators have been found is to construct the
operators level by level in the variable $m+n$. One operator, \eg $L_{30}$,
may be used to step up to the next level and the operators $L_{20}$ and
$L_{02}$ may be used to complete the spin multiplets. With the help of these
three operators all the others can be generated. Identities between the
operators can be found by reducing all operators to a standard form by use of
the modified Heisenberg commutation relation (\ref{modcom}).

\section{Other formulations of the algebra}
The expressions for the observables $L_{mn}$ given above refer to the
representation of the algebra in the Calogero model. I will now scetch how the
algebra can be expressed in a more general way \cite{Isakov 96b}. To do so let
us consider the the adjoint representation, with each element expressed as a
product of $a$ and $a^\dagger$. The observables $L_{mn}$ then act on products
of $a$ and
$a^\dagger$ and map them into new products. We may think of the following
commutators as the fundamental ones,
\be
&[L_{m0},a^\dagger]=0  &[L_{m0},a]=-ma^{\dagger (m-1)} \nonumber \\
&[L_{0n},a^\dagger]=na^{n-1}\;\;\;\;&[L_{m0},a]=0
\label{baserel}
\ee 
From these expressions the action of a general element $L_{mn}$, which is
defined by repeated commutations of $L_{m0}$ and $L_{0n}$, can be deduced. The
only additional input needed is that an observable $L_{mn}$, when acting on a
product of $a$ and $a^\dagger$, gives a sum of products where the observable
acts on each element in the product, \eg
\be
[L_{mn}, a a^\dagger a ...]= [L_{mn}, a] a^\dagger a+ a  [L_{mn},a^\dagger]a+ a
a^\dagger [L_{mn},a] + ...
\label{Leib}
\ee
This trivially follows from the fact that the observable act on the
product by commutation. 

Note that in this construction we do not need to establish the explicit
expressions for the observables in terms of $a$ and $a^\dagger$. Therefore the
question of reordering of operators does not enter. Also note that the result of
acting with an observable on a product of $a$ and $a^\dagger$ is an expression
with a unique ordering of operators. Let us therefore forget about the Calogero
model and {\em define} the algebra of observables $L_{mn}$ by the relations
(\ref{baserel}) and (\ref{Leib}). We assume no commutation rule for $a$ and
$a^\dagger$, so different orderings have to be considered as distinct. We note
that (\ref{Leib}) has the form of Leibniz rule so the observables $L_{mn}$ can
be interpreted as differential operators acting on the products (the
non-commutative ring) generated by $a$ and $a^\dagger$. Clearly these
differential operators define an infinite-dimensional algebra.

The claim is now that the algebra defined in this more abstract way is the same
as the one defined by use of the operators of the Calogero model. This may
seem strange since we have not taken into account the identities which
earlier created problems for an explicit construction of the algebra. But the
identities are in fact there, although in a somewhat different form. An element
$I_{mn}=L_{mn}-L_{mn}'$ , in the new definition of the algebra, is identified
with zero if it commutes with all products of $a$ and $a^\dagger$. A sufficient
condition for this to happen is that
\be
[I_{mn},a]=[I_{mn},a^\dagger]=0
\ee
Therefore the identities are automatically taken care of when the
operators $L_{mn}$ are defined by their action on the products of $a$ and
$a^\dagger$. 

To see the correspondance with the algebra of the Calogero model more
explicitly, we note that the defining relations (\ref{baserel}) there can be
represented in the following form
\be
&[\sum_i a_i^{\dagger m},a_j^{\dagger}] = 0 \;\;\;\;
&[\sum_i a_i^{\dagger m},a_j]=-ma_j^{\dagger (m-1)} \nonumber \\ 
\ &[\sum_i a_i^n ,a_j^{\dagger}] = na_j^{n-1}\;\;\;\;&[\sum_i a_i^n,a_j]=0
\ee
Note that even if there is no summation over the particle index $j$, these
relations are independent of the statistics parameter $\nu$. This implies that
any observable $L_{mn}$ when commuted with a product of $a_j$ and $a_j^\dagger$
gives a sum of products of these operators -- only referring to particle $j$ --
which is independent of $\nu$. In particular this is true for the Calogero
Hamiltonian (see (\ref{Hcom})), and this fact has already been exploited in
\cite{Brink 92} to construct eigenstates and eigenvalues of the Hamiltonian.
For products of $a_j$ and $a_j^\dagger$, with no sum over $j$, a re-ordedring
of operators by use of the modified Heisenberg commutation relations will give
terms which depends on
$\nu$. It is only after the summation over the particle index that the
$\nu$-independent identities appear. The correspondence between the operators
$a$ and $a^\dagger$ in (\ref{baserel}) and the un-summed Calogero operators
therefore gives an explanation why these identities should not be implemented.

As already mentioned we may view the observables $L_{mn}$ as linear
differential operators acting on products of $a$ and $a^\dagger$. A general
element of the algebra can be written in the form 
\be
L=P(a,a^\dagger)\frac{\partial}{\partial
a^\dagger}+Q(a,a^\dagger)\frac{\partial}{\partial a} 
\ee
where $P$ and $Q$ are polynomials in $a$ and $a^\dagger$. This formulation of
the observables in fact introduces a simplification in the construction of the
algebra. Commutators can be calculated by the use of these expressions, and
there is no ambiguity in the resulting expressions due to identities between
reordered products of $a$ and $a^\dagger$. However, I will now discuss another
representation, in terms of matrices, which gives an even simpler way to do the
construction.

\section{The matrix model}
When discussing the algebra $W_{1+\infty}$ of bosons and fermions, the
presence of higher order terms in $\hbar$ was pointed out. For the
algebra of observables discussed here such higher order terms are not
present. In the construction of the algebra by use of the operators
of the Calogero model this can be understood by the way the commutators are
evaluated, namely by applying the (modified) Heisenberg commutation rule only
once in the evaluation of a commutator. It was not used repeatedly to bring the
expression into a standard form; instead expressions with different ordering of
operators were accepted as different elements of the algebra.

The lack of higher order terms in $\hbar$ indicates that the same
algebra is present also at the classical level. This is in fact the
case, and the defining relations (\ref{modcom}) are clearly satisfied
for a classical $N$-particle system if the commutators are replaced by
Poisson brackets, and the quantum variables $a$ and $a^\dagger$ are
replaced by their classical counterparts (or by $x$ and $p$). But
the algebra is not represented fathfully in this way. Since $x$ and
$p$ commute, there is no distinction between expressions differing
by reordering these variables. If the algebra should be represented
faithfully at the classical level we have to consider a system
where the phase space variables do not commute at the classical
level.

A classical system where $x$ and $p$ do not commute 
may seem rather unconventional. However, the matrix models are
exactly of this type. In these models the coordinate $x$ as well as
the momentum $p$ are matrix-valued phase-space variables. Even if
each matrix element of $x$ commutes with each matrix element of
$p$, they do not commute as matrices. For finite $N$ there will exist certain
relations between reordered (matrix) products of $x$ and $p$, but in the limit
$N\rightarrow \infty$ there will be no relations surviving between reordered
products.

The fundamental Poisson brackets for the matrices $x$ and $p$ are
\be
\{x_{ij},p_{kl}\}_{pb}=\delta_{il}\delta_{jk}
\ee
and for general functions of $x$ and $p$ they get the form
\be
\left\{ {C,C'} \right\}_{pb}=\sum\limits_{ij} {\left[ {{{\partial C}
\over {\partial x_{ij}}}{{\partial C'} \over {\partial p_{ji}}}-{{\partial
C} \over {\partial p_{ji}}}{{\partial C'} \over {\partial x_{ij}}}} \right]}
\label{pb}
\ee
By use of these bracket relations it is straight forward to demonstrate that
the defining relations of the algebra (\ref{modcom}) are satistied. Expressed
in terms of $x$ and $p$ (rather than $a$ and $a^\dagger$) they have the form
\be
\{Tr\, p^m,p\}_{pb} &=& 0 \;\;\;\;
\{Tr\, p^m,x\}_{pb}=-mx^{m-1} \nonumber \\ 
\ \{Tr\, x^n ,p\}_{pb} &=& nx^{n-1}\;\;\;\;\{Tr\, x^n,x\}_{pb}=0
\ee
Comparing with the expressions of the Calogero model, we note that the products
with no sum over the particle index $j$ in the present case correspond to the
matrix products of $x$ and $p$ (without taking the trace) whereas the products
with summation over the particle index correspond to the trace of the matrix
products. The elements of the algebra then correspond to such traced matrix
products.

The matrix formulation now can be used to construct the algebra step by step in
the same way as discussed for the Calogero model. The Poisson brackets are
evaluated by use of (\ref{pb}), and in the evaluation the order of the
variables of the matrix products have to be respected. This matrix
representation is simpler for two reasons. Firstly, the cyclic property of the
trace means that there are fewer products of $x$ and $p$ to care about.
Secondly, there are no additional (hidden) identities between products with
different ordering of $x$ and $p$ to worry about. It is of interest to note
that no matrix multiplication has actually to be performed. The evaluation of
Poisson brackets can be formalized in a diagrammatic form where a matrix
product is represented by an ordered string of two different objects ($x$ and
$p$). The untraced product is an open string and the traced product is a closed
string. The Poisson bracket is implemented as a rule for merging two of these
strings.    

To construct the algebra in the matrix representation, the rules for how to
evaluate Poisson brackets on a phase space of non-commuting $x$ and $p$ is all
we need. It is nevertheless of interest to be more specific about the matrix
model in order to see the the correspondence with the original system of $N$
identical particles in one dimension \cite{Polychronakos 96}. Let me then
assume $x$ and $p$ both to be Hermitian
$N\times N$ matrices with a free particle Hamiltonian of the form
\be
H=\half Tr\,p^2 
\ee
The time evolution in matrix form is simply
\be
x=x_0+vt
\ee
with $x_0$ and $v$ as constant matrices. The position and velocity do not have
to commute as matrices, but the commutator is a constant of motion
\be
K=i[x,\dot x]=i[x_0,v]
\ee

The Hamiltonian as well as all other elements of the algebra of observables,
which can be written as the trace of products of $x$ and $p$, are invariant
under $SU(N)$ transformations of the form
\be
x\rightarrow U x U^\dagger\;\;\;\;p\rightarrow U p U^\dagger
\ee
In particular this is the case for the invariant coordinates $x^{(k)}=Tr(x^k)$
which determine the eigenvalues of the matrix $x$. These eigenvalues we may
view as the coordinates of a one-dimensional $N$-particle system. The $SU(N)$
group may be regarded as an extension of the permutation group for this
$N$-particle system and it allows the positions of the $N$ particles to be
continuously permuted. This continuous transformation involves the
non-invariant parts of $x$ and the the extension of the permutation group to
$SU(N)$ can be seen as a compensation for the fact that more degrees of freedom
are present in the matrix model than in the $N$-particle system.

A separation of the $SU(N)$ invariant and the non-invariant coordinates can be
made explicit by a (time-dependent) diagonalization of $x$,
\be
x=U^\dagger x_d U
\ee
Here $x_d$ is a diagonal matrix which depends on the (invariant) eigenvalues,
while
$U$ depends on a set of (generalized) angular variables which are non-invariant
under the $SU(N)$ transformations. Expressed in terms of $x_d$ and $U$ the
Hamiltonian gets the form
\be
H=\half Tr(\dot x_d^2 + J^2)
\ee
where
\be
J=[x_d,\dot U U^\dagger]
\ee
We note that if the angular velocities $\dot U U^\dagger$ can be traded with a
set of conserved quantities (generalized angular momenta), one may be able to
write $H$ as a function only of the variables $x_d$, their time-derivative and
these conserved quantities. The Hamiltonian then describes a set of $N$
interacting particles with
$\half Tr(J^2)$ as the potential. This reduction of the number of variables
clearly is analogous to the reduction to the radial variable for a rotationally
symmetric one-body problem.

An explicit construction of this type can be given \cite{Polychronakos 96}. Let
us choose a basis where the coordinate $x_0$ at time $t=0$ is diagonal and give
the system an initial velocity $v$ with non-diagonal matrix elements
\be
v_{ij}= -i\frac{l}{x_{0i}-x_{0j}},\;\;\;\; i\neq j 
\ee
with $x_{0i}$ as the eigenvalues of $x_0$. (The diagonal matrix elements of $v$
are arbitrary.) The conserved quantity
$K$ then has non-diagonal matrix elements which all are equal,
\be
K_{ij}=l,\;\;\;\; i\neq j 
\ee
while the diagonal matrix elements necessarily vanish. We have the following
relation between $K$ and $J$,
\be
UKU^\dagger=i[x_d,J]
\ee
which in component form gives
\be
l\sum_{m\neq n} U_{im}U_{jn}^*=i(x_i-x_j)J_{ij}
\label{UKU}
\ee
with $x_i$ as the eigenvalues of $x$. From the diagonal case $i=j$ and the
unitarity of $U$ we deduce
\be
|\sum_m U_{im}|^2=1
\ee
which, when re-inserted in (\ref{UKU}) gives
\be
J_{ij}=-i\,l\frac{\alpha_i\alpha_j^*}{x_i-x_j},\;\;\;\; i\neq j
\ee
where $\alpha_i=\sum_m U_{im}$ is a phase factor. This gives the following
expression for the Hamiltonian
\be
H= \half (\sum_i \dot x_i^2+\sum_{i\neq j}\frac{l^2}{(x_i-x_j)^2})
\ee
which is the same as that of  a (classical) system of $N$ identical particles
interacting through a $1/x^2$ two-body potential. 

The similarity between the interacting $N$-particle system and a free system now
becomes clearer. The system interacting with a two-body $1/x^2$ potential can be
derived from a free system with more degrees of freedom, namely the $N\times N$
matrix model, and the potential arises in a similar way as a sentrifugal
potential of a free particle in 2 or 3 dimensions when this is described in
radial coordinates. The correspondence between the classical systems discussed
here carries over to the quantized systems, but then with restrictions on the
possible values of $l$ due to the compactness of the $SU(N)$ group. 

\section{Concluding remarks}
The purpose of the work which has been reviewed here is to examine
generalizations of quantum statistics from the point of view of Heisenberg
quantization. This approach is interpreted as a problem of finding
representions of a fundamental algebra of observables for a system of $N$
identical particles. For 2 particles in one dimension the algebra is identified
as $su(1,1)$, and the representations are readily found. There are
generalizations of the Bose and Fermi cases which can be expressed in terms of
a statistical $1/x^2$ interaction. For the
$N$ particle system the statistical interaction is assumed to be a two-body
interaction of the same form. In this way one is lead to consider the Calogero
model as desribing a system with generalized statistics.

Much of the discussion I have given concerns the question of whether a
fundamental algebra of observables can be identified for the $N$-particle
system, an algebra which replaces $su(1,1)$ for two particles. Since the
Calogero model is assumed to describe a system with generalized statistics, the
operators of this model can be used to construct the algebra. A main result is
that such an algebra, which is independent of the (statistics) parameter, can
indeed be constructed. The algebra is infinite-dimensional, and even if a
closed formulation of the algebra has not been found, a method for constructing
a basis for the algebra step by step can be given.  An interesting point which
has been emphasized is that a more general formulation of the algebra can be
given than the one given in terms of operators of the Calogero model. This
general formulation can be used to demonstrate the presence of the same algebra
also in other systems, in particular in the Hermitian matrix model. 

There remain several unsolved problems concerning the algebra of observables
which has been discussed here. A main mathematical problem is to identify more
fully the structure of the algebra. Another question concerns the possibility
of finding other (physically) interesting representations. In addition to these
questions about the mathematical structure of the algebra, one has the
important question of whether there are physical systems where this structure
appear in a natural way. When the algebra of the 2-particle system was
originally introduced it was applied to the system of vortices in a superfluid
helium film \cite{Leinaas 88}. It may be questionable  whether such vortices
are good candidates for the application of ideas of (one-dimensional)
statistics, but there exist other vortex-like excitations which may be. Anyons
in a strong magnetic field have been demonstrated to satisfy this type of
statistics \cite{Hansson 92}, and the correspondence to the Laughlin
quasi-particle excitations suggest that the algebra discussed here may be
relevant for descriptions of the quantum Hall system. However, my discussion 
has aimed mainly at demonstrating how, in the framework of Heisenberg
quantization, the algebra of observables for a system of identical particles in
one dimension appears as unique mathematical structure. The other interesting
questions about further investigations and about possible applications I will
have to leave as open questions for future research.


\end{document}